# CONTEXT-CAPTURE MULTI-VALUED DECISION FUSION WITH FAULT TOLERANT CAPABILITY FOR WIRELESS SENSOR NETWORKS

Jun Wu[1] and Shigeru Shimamoto[2]

[12] Graduate School of Global Information and Telecommunication Studies, Waseda University, Tokyo, Japan
junwu@akane.waseda.jp, Shima@waseda.jp

## ABSTRACT

*Wireless sensor networks (WSNs) are usually utilized to perform decision fusion of event detection. Current decision fusion schemes are based on binary valued decision and do not consider bursty context-capture. However, bursty context and multi-valued data are important characteristics of WSNs. One on hand, the local decisions from sensors usually have bursty and contextual characteristics. Fusion center must capture the bursty context information from the sensors. On the other hand, in practice, many applications need to process multi-valued data, such as temperature and reflection level used for lightening prediction. To address these challenges, the Markov modulated Poisson process (MMPP) and multi-valued logic are introduced into WSNs to perform context-capture multi-valued decision fusion. The overall decision fusion is decomposed into two parts. The first part is the context-capture model for WSNs using superposition MMPP. Through this procedure, the fusion center has a higher probability to get useful local decisions from sensors. The second one is focused on multi-valued decision fusion. Fault detection can also be performed based on MVL. Once the fusion center detects the faulty nodes, all their local decisions are removed from the computation of the likelihood ratios. Finally, we evaluate the capability of context-capture and fault tolerant. The result supports the usefulness of our scheme.*

## KEYWORDS

*Wireless Sensor Networks, Decision Fusion, Context, Fault-Tolerant, Multi-Valued Logic*

## 1. INTRODUCTION

Wireless sensor networks (WSNs) have become a technology for the new millennium with the endless for applications ranging from civilian to military. WSNs are usually utilized to perform decision fusion of event detection. Sensors often make independent local decisions based on their local observations and transmit these decisions to a common fusion center (FC). The FC combines the signals received from the sensors based on some fusion rule to generate the final decision [1]. The fusion center may be a central decision unit, or, in clustered-based WSN, may simply be a cluster head.

Although there have been a lot of works studied decision fusion for WSNs, two important facts have not been considered in the existing schemes: bursty context and multi-valued characteristics of the detected events.

On one hand, successful event detection is typical contextual. For fusion center, the local decisions from the sensors are context information. There is a need for a fusion center that is able to capture the local decisions from the sensors, which is context decision information over different kinds of event sources. For example, in an intelligent home network [2], some sensors are used to measure the temperature, while some other sensors are used to monitor the humidity in the house. Obviously, the change rates of the temperature and the humidity are different.

Hence, different kinds of events usually happen with different rate, which means that they have bursty characteristic. Nowadays Poisson process is usually used for model the event in WSNs [3], [4]. However, Poisson process is not able to capture the burstiness context of the events. Therefore, how to capture the bustiness context is very important for WSNs.

On the other hand, in current decision making schemes for WSNs, sensors transmit binary decisions to the fusion center at which they are combined to yield multiclass decisions. In practice, many applications need to process multi-valued data, such as temperature and reflection level used for lightening prediction [5]. Binary decisions of sensors cannot resolve this problem. Moreover, because of the faults of sensors or the noise in the wireless channels, the fusion center receives a vector of potentially distorted decision information from sensors. In other words, in fusion center, the information used to make global decision is usually incomplete or uncertain. The binary decision cannot provide accurate decision according to the incomplete or uncertain information. Hence, binary decision fusion cannot satisfy efficiently the applications of WSNs. Also, it has been indicated that, in a WSN, fault tolerance capability is critical since sensors can be damaged, blocked or run out of battery energy [6], [7]. Multi-valued logic is a great tool to provide fault detection [8]. Hence, based on the above observations, we consider using multi-valued logic to perform the decision fusion in WSNs.

To address above two challenges, we proposed a context-capture multi-valued decision fusion scheme for WSNs. The proposed scheme realizes the context-capture and multi-valued decision fusion based on Markov modulated Poisson process (MMPP) and multi-valued logic (MVL), respectively. Also, the proposed scheme can provide fault tolerant capability for decision fusion in WSNs. The rest of this paper is organized as follows. In Sect. 2, we describe related works in the area of decision fusion in WSNs. In Sect. 3, we present the architecture of decision fusion and the preliminaries in our scheme. In Sect. 4, we discuss the context-capture principle based on MMPP. We present multi-valued decision fusion based on MVL in Sect. 5. Then, the overall procedure for the proposed context-capture multi-valued decision fusion is illustrated in Sect. 6. In Sect. 7, we evaluate the capability of context-capture and fault tolerant of our scheme. Finally, we conclude the paper in Sect. 8.

## 2. Related Works

There have been several studies focused on the decision fusion of WSNs. In canonical sequential testing or decision fusion [9]-[11], the final decision is made when likelihood ratios are greater or less than the thresholds set. Thus the results can satisfy the requirements of error probability or some cost. In the works of [12], [13], the sensors send local binary decisions to the fusion center. Then the local decisions data are combined to get multiclass decisions. In the schemes in [14]–[16], the adaptive approaches are all based on estimating the local sensor error probabilities. Then these estimates can be applied in the Chair-Varshney fusion rule. Unlike the conventional approach that employs the Chair-Varshney fusion rule1 that assumes no faults, the fault-tolerant fusion rule provides enough distance between the decision regions corresponding to different hypotheses by a careful design and exploitation of a code matrix.

The existing decision fusion schemes in WSNs mainly focus on the optimal fusion rule. However, there are two important facts have not been considered in the existing schemes: bursty context and multi-valued characteristics of the detected events.

## 3. Preliminaries

### 3.1. Architecture and Difficulties

The architecture of the decision fusion is shown in Fig. 1. The sensors process their respective observations { $c_1, c_2, \ldots, c_n$ } independently by forming local decisions {$d_1, d_2, \ldots, d_n$}. Then the sensors send the local decisions to the fusion center in order to perform decision combining. Due to the noisy channels and interference, the fusion center receives a vector of potentially distorted decisions $\{d_1', d_2', \ldots, d_n'\}$. It is preferable that the fusion center makes error global decision based on the distorted information from the sensors.

As mentioned in Sect. 1, the rates of the local decisions from different sensors are different, so the fusion center must be able to capture the bursty contexts from sensors. Moreover, existing binary decision cannot describe efficiently the event. Hence, WSNs need context-capture multi-valued decision fusion. Meanwhile, during the design of context-capture multi-valued decision fusion, it is still very necessary to increase the fault tolerance capability.

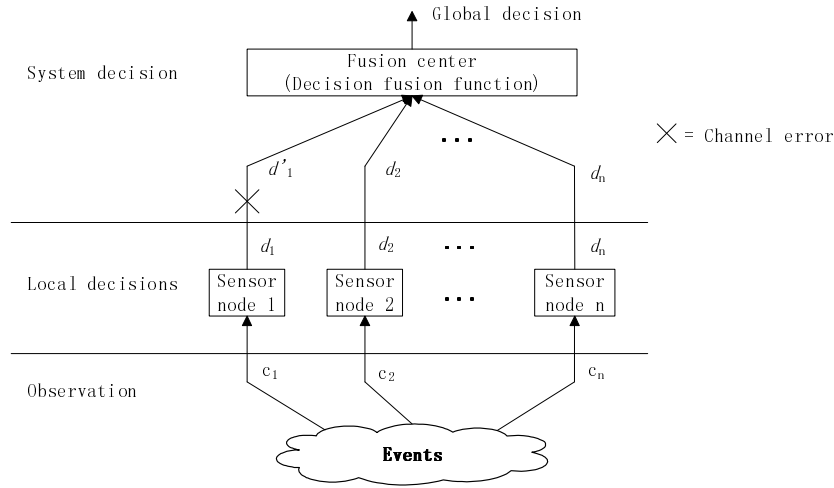

Figure 1. Architecture of decision fusion in WSNs

## 3.2. Markov Modulated Poisson Process

Markov Modulated Poisson Process (MMPP) [17] is an n-state Markov chain that has different rates in each state and with more parameters compared with Poisson process. In other word, MMPP is a doubly stochastic Poisson process in which a Markov chain governs the transition of the process between phases, and at each phase (or state) the number of arrival in a time frame is determined by Poisson process. MMPP is often used in queuing theory [18]-[20], but it is rare in decision fusion. MMPP can model a Poisson process whose rate parameter varies according to a Markov process. In other word, MMPP is particularly useful in modelling time-varying intensity rate processes. Therefore, MMPP is a flexible model for point processes whose event rates vary among different levels at irregular intervals. In this paper, periodic event contexts are modelled via a time-vary Poisson process model. However, bursty event contexts are modelled through a MMPP.

## 3.3. Multi-Valued Logic

Multiple-valued logics (MVL) can explicitly represent uncertainty and disagreement. Thus they can be applied to the modeling of complex behaviour of real system, especially the exploratory modelling used in the early stage of requirements engineering and architectural design, to verify properties of models that contain uncertainty or disagreement. MVL is a proven technology which has been investigated for many years [21], [22]. Besides reduction in chip area as well as fault tolerance, MVL offers other benefits, such as potential for speech recognition [23]. Also, MVL can be applied in the area of decision making [24].

In this paper, we introduce MVL into the decision fusion in WSNs. Thus our scheme can not only satisfy the multi-valued characteristics of detected events, but also provide efficient fault tolerant capability for WSNs.

### 3.4. Basic Principle

Based on the analysis of study difficulties and possible solutions, the basic principle of context-capture multi-valued decision fusion for WSNs is illustrated in Fig. 2.

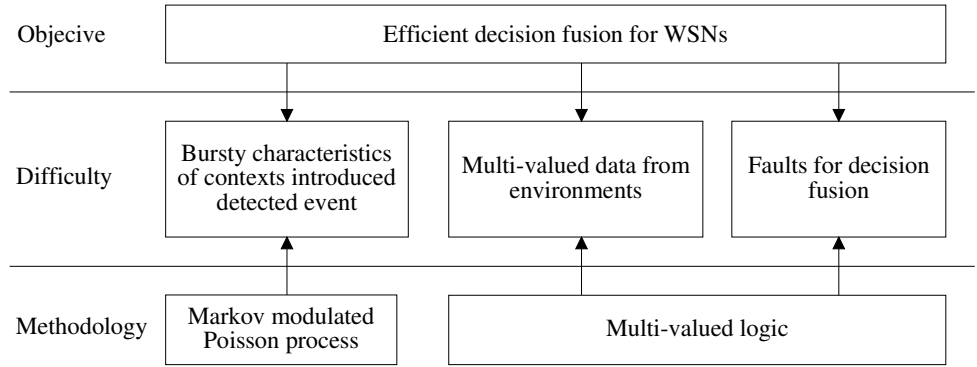

Figure 2. Basic principle

## 43. CONTEXT-CAPTURE BASED ON SUPERPOSITION MMPP

In this section, we present the principle of the context-capture in our scheme. We use the two state Markov process governs the bursty contexts of the abnormal events' presence or absence, and that periodic event contexts are modelled via a time-vary Poisson process model.

### 4.1. Normal Context-Capture

To model the periodic, predictable portion of contexts corresponding to normal event, we use nonhomogeneous Poisson process with a particular parameterization of the rate $\lambda$, which is derived from that of the model in [25]. The probability mass function of the Poisson distribution is given by

$$P(x;\lambda) = e^{-\lambda}\lambda^x / x!, \quad (x = 0,1,2,...) \tag{1}$$

where the parameter $\lambda$ represents the rate, or average number of occurrences in a fixed time interval. For nonhomogeneous Poisson distribution, $\lambda$ is a function of time, namely $\lambda(t)$. The degree of heterogeneity depends on the function $\lambda(t)$.

Let $Y(t)$ is a measurement of the number of contexts engaged in some event over the time interval [$t$, $t$+1]. We assume $Y(t)$ consists of two elements, which is given by

$$Y(t) = Y_N(t) + Y_C(t), \quad Y(t) \ge 0 \tag{2}$$

where $Y_N(t)$ is the number of contexts attributed to the normal building occupancy, and $Y_C(t)$ is the change in number of occurrences which is attributed to an event at time t (positive or negative); the nonnegativity condition indicates that we cannot observe fewer than zero counts.

## 4.2. Bursty Context-Capture

Markov Modulated Poisson Process (MMPP) [17] is an n-state Markov chain that has different rates in each state and with more parameters compared with Poisson process. MMPP is known as the Switched Poisson Process (SPP), which works as a very versatile tool for the modelling of variable context-aware. MMPP is often used in queuing theory [18]-[20], but it is rare in decision fusion. We use superposition of two-state Markovian sources [20] to model the multi-valued local decision contexts from sensors. The superposition is actually a MMPP which is a special case of the Markovian Arrival Process (MAP) [26]. The Markov process introduces bursts of context data associated with episodes of abnormal events.

The model consists of a superposition of $N$ two-state Markov models. For simply presentation, the $k$-th MMPP ($1 \leq k \leq N$) is denoted by MMPP$^{(k)}$. An two-state MMPP$^{(k)}$ can be formulized by the infinitesimal generator, $G_k$, and rate matrix $R_k$ as

$$G_k = \begin{bmatrix} -\delta_{1k} & \delta_{1k} \\ \delta_{2k} & -\delta_{2k} \end{bmatrix} \tag{3}$$

$$R_k = \begin{bmatrix} r_{1k} & 0 \\ 0 & r_{2k} \end{bmatrix} \tag{4}$$

where transition rate from state 1 to 2 of the MMPP$^{(k)}$ is $\delta_{1k}$, and the rate out of state 2 to 1 is $\delta_{2k}$. Meanwhile, $r_{1k}$ and $r_{2k}$ are context renew rates in case of the MMPP$^{(k)}$ is in state 1 and 2, respectively. The superposition of these gives rise to a new with $2N$ sates and its parameters, $G$ and $R$, can be calculated as

$$(G, R) = (\bigoplus_{k=1}^{N} G_k, \bigoplus_{k=1}^{N} R_k) \tag{5}$$

where $\oplus$ denotes the Kronecher sum [17]. We represent the steady-state probabilities by vector

$$\theta = [\theta_{1k}, \theta_{2k}] \tag{6}$$

The MMPP can be constructed from the superposition of $N$ 2-state MMPPs. Then the autocovariance and marginal distribution of the rate process should be computed. The autocovariance of the rate process of a MMPP with $2N$ states can be distribution of the rate process. The autocovariance of the rate process of a MMPP with $2N$ states can be described as a weighted sum of $N$ exponentials, i.e.,

$$\sigma(t) = \sum_{j=1}^{N} \sigma_j(t) = \sum_{j=1}^{N} \alpha_k e^{-\beta_j t} \tag{7}$$

The autocovariance of $k$-th 2-state MMPP can be computed as follow

$$\sigma_j(t) = g_k^2 \theta_{1k}(1-\theta_{1k})e^{-(\delta_{1k}+\delta_{2k})t}, \quad k, j = 1, 2, ..., N \tag{8}$$

The difference between the Poisson rates of the two states is $g_j$, which can be computed by

$$g_j = \delta_{2k} - \delta_{1k} \tag{9}$$

where $\alpha_j$ and $\beta_j$ is given by

$$\alpha_j = g_k^2 \theta_{1k}(1-\theta_{1k}) \ , \ k, j = 1, 2, ..., N \tag{10}$$

$$\beta_j = \delta_{1k} + \delta_{2k} \ , \ k, j = 1, 2, ..., N \tag{11}$$

Thus, the constraints imposed by the autocovariance matching on the parameters of the superposed model can be computed by above equations.

Assume index $k, j = 1, 2, ..., N$ represent the states of the generic MMPP and let indexes $(h, k)$, $k = 1, 2, ..., N$ and $h \in \{1, 2\}$, represent state $h$ of $k$-th 2-state MMPP of the superposed MMPP. There is a correspondence between each state of the generic MMPP and a set of $N$ states of the superposed MMPP. This set can be computed by

$$M_i = \left\{ (h, k) : k = 1, 2, ..., N; h = 2 - \text{mod}\left( \left\lceil \frac{i}{2^{N-k}} \right\rceil, 2 \right) \right\} \tag{12}$$

where mod($x$,2) represents $x$ modulus 2 and $\lceil y \rceil$ represents the lowest integer greater than $y$.

Using this definition, the arrival rates and steady-state probabilities of the generic MMPP can be obtained from those of the superposed MMPP by

$$r_i = \sum_{(h,k) \in M_i} \delta_{hk} \ , \ i = 1, 2, ..., N \tag{13}$$

$$\pi_i = \prod_{(h,k) \in M_i} \theta_{hk} \ , \ i = 1, 2, ..., N \tag{14}$$

where is the steady-state probabilities of the generic MMPP.

From Eq. (12), the mapping relations can be defined by (12). Assume $r_\Delta$ denotes the arrival rate of the generic MMPP. Then $N$ arrival rate differences of the superposed MMPP can be computed as follows

$$r_i = r_\Delta + \sum_{k=1}^{N} \left( d_k \left( 1 - \text{mod}\left( \left\lceil \frac{i}{2^{N-k}} \right\rceil, 2 \right) \right) \right) \tag{15}$$

where $i = 1, 2, ..., 2^N$.

According to Eq. (10), because the autocovariance imposes only the arrival rate, differences will give an additional degree of freedom for matching the distribution. There will be one (or more) solutions for mapping the generic MMPP into a superposed MMPP only if matrices $G_i$ and $R_i$ can be determined from Eq. (3).

## 5. MULTI-VALUED DECISION FUSION

In this section, we present the multi-valued decision fusion. We adopt a multi-valued fault detection rule for WSNs, then propose a scheme to remove the all corresponding fault local decisions to perform the fault tolerant.

### 5.1. Fault Detection based on MLV

Let $D=(d_1, d_2,\ldots, d_n)$ be a $g$-valued $n$-variable input vector, where $x_i \in L$, $L \in \{0,1,2,\ldots, g\text{-}1\}$, $i \in \{1,2,\cdots,n\}$. Also, let $f(x)$ be a $g$-valued $n$-variable logic function mapping from $L_n$ to $L_1$, and $(u_1,u_2,\ldots,u_n)$ be the $g$-ary expansion of $U$ with $u_1$ the least significant position. We shall write $f(x)$ to denote the value of $f$ when $d_i=u_i$ ($1 \leq i \leq n$). The spectrum $m_0,m_1,\ldots,m_q (q=g^{n-1})$ of $f(d)$ is defined by [27].

$$s_w = \sum_{u=0}^{q} \overline{\exp(-\sqrt{-1}\frac{2\pi}{g})^{\sum_{i=1}^{n} w_i u_i}} y(u) \tag{16}$$

From Eq. (16), y(w) can be computed by

$$y(w) = \frac{1}{g^n}\sum_{u=0}^{q} t_w(u)s_u \tag{17}$$

Then, a matrix notation for convenience is adopted. All vectors are column vectors. Let $Y$ be the vector whose $u$-th element is $y(u)$ and $S$ be the vector whose $w$-th element is $s_w$. $T_g^n$ is $g^n \times g^n$ matrix whose element in the $w$-th row and $u$-th column is $t_w(u)$.

$$[Y] = \frac{1}{g^n}[T_g^n \mid S] \tag{18}$$

where $t_w(u)$ is computed as follow

$$t_w(u) = \exp(-\sqrt{-1}\frac{2\pi}{g})^{\sum_{i=1}^{n} w_i u_i} \tag{19}$$

A single input stuck-at fault is a fault where one of decision $(d_1, d_2,\ldots, d_n)$ are stuck a certain value, which is independent of local detection. Based on the method of analysis in [8], a single fault of primary $(d_1, d_2,\ldots, d_n)$ in the multiple valued logic is syndrome testable if and only if

$$\sum_{k=1}^{g-1} t_{ki}^1 s_0^k \neq 0 \tag{20}$$

### 5.2. Multi-Valued Decision Fusion with Fault Tolerant Capability

Assume there are $n$ sensors in the system, and every sensor can make a local decision $d \in (0, 1, 2, \ldots, g\text{-}1)$. The goal of decision fusion is to determine the true hypothesis from the $g$ hypotheses $H_i$ ($i$=0, 1, 2,… , $g$-1). The fusion center uses the local decision $d$ as its observations. In a decision-making cycle, assume there $k_i$ local decisions are sent to fusion center from the $i$-th sensor. Fusion center make the global decision based on the $K$ local decision contexts, where $k=n·k_1+n·k_2+...+n·k_n$.

For the $j$-th sensor, assume $R(i)$ is the likelihood ratio. Then the local fusion rule is

$$d_j = \begin{cases} g-1, & if \ \ R_{ij} < 0, \forall i = 0,1,...,g-2 \\ k, & otherwise, k = \arg\max_{0 \le i \le g-2}\{R_{ij}\} \end{cases} \qquad (21)$$

The global fusion is based on minimum error probability rule. Therefore, the log likelihood is defined as

$$L_i(D) = \log \frac{P(H_i / d)}{P(H_{g-1} / d)} \qquad (22)$$

where $i$=0, 2, …, $g$-2, and $d$ is group of decision context sent to fusion center. Then, at fusion center, the fusion rule can be got as follow

$$z = \begin{cases} g-1, & if \ \ L_i < 0, \forall i = 0,1,...,g-1 \\ k, & otherwise, k = \arg\max_{0 \le i \le g-2}\{L_i\} \end{cases} \qquad (23)$$

The fusion center also makes the decision based on the log likelihood. If faulty nodes are found, the corresponding local decisions are removed from the computation of the likelihood ratios, which are used by fusion center to make final decisions. Therefore, real-time decision fusion with faulty sensor identification may be applied in WSNs with a very large number of nodes.

## 6. PROCEDURE OF CONTEXT-CAPTURE MULTI-VALUED DECISION FUSION

The overall procedure of the security assessment is illustrated in Fig. 3. As shown in Fig. 3, the procedure of local multi-valued decision making is perform in sensors, while the procedure of decision fusion is perform in fusion center.

Every sensor makes the local decisions based their observations. The local decision is made in a multi-valued form. After finishing the local decision, the sensors send the local multi-valued decision to fusion center. Upon receiving the local decisions from different sensors, the fusion center capture the decision contexts based on superstition MMPP, because the rate of the decisions from different sensors is different. Then the fusion center analyzes every local decision from the sensors. If a fault is detected, the fault decision will be removed from the computation of the likelihood ratios, which are utilized to make final decisions regarding which hypothesis is true. Finally, the fusion center makes the final decisions based on the true local decisions.

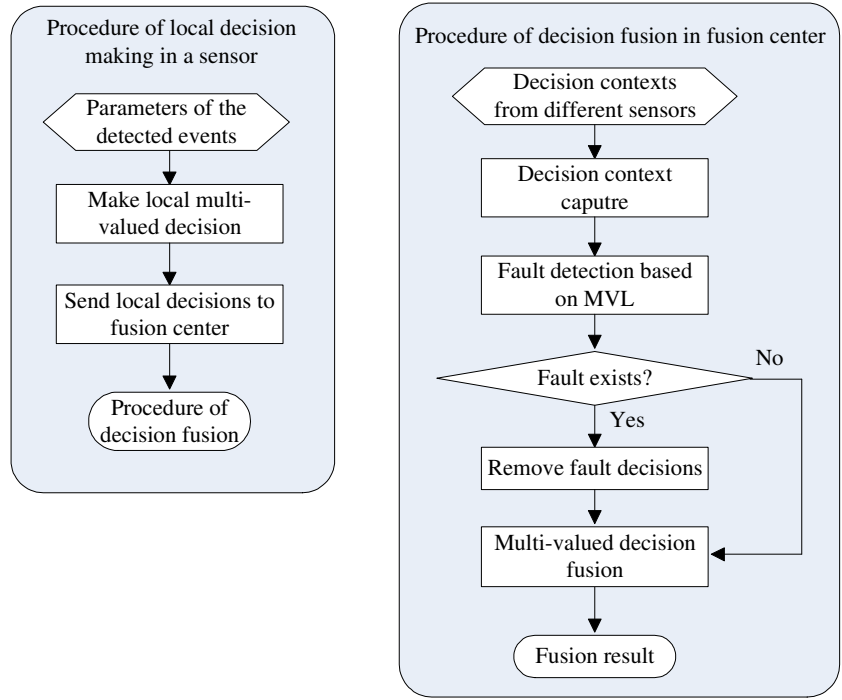

Figure 3. Procedure of context-capture multi-valued decision fusion

# 7. SIMULATION AND EVALUATION

In this section, we evaluate the performance of the proposed scheme.

## 7.1. Context-Capture

For evaluating the context-capture capability, we simulate a network model which consists of two sensors and a fusion center. The rates of local decision the two sensors are different, so the system has bursty characteristic. Because here we focus on evaluating the context-capture capability of our scheme, we do not consider channel error in this case. The channel error will be considered when we evaluate the fault tolerance. In every sensor, the local decision is made according to an interrupted Poisson process because the "off" state prevents entity generation. The on-off modulated Markov sources for local decision are independent. Their behaviors depend on the rate of the Poisson process when the Markov chain is in the "on" state. Then the fusion center capture the context based on the superposition MMPP in Sect. 3.

Let $\tau$ denotes average time between On-Off points, and $r$ denotes rate. For simplicity of the presentation, we assume the system is in the case with the parameters as follows: $\tau_1$=30, $r_1$=1/15 for sensor-1, and $\tau_2$=50, $r_2$=1/10 for sensor-2.

Next, we observe the local decision in two sensors and the context-capture in fusion center. The decision context generations in sensor-1 and sensor-2 are shown in Fig. 4 and Fig. 5. A line denotes that a local decision is made in the sensor according to the observations in the environments. We evaluate the context-capture in fusion center. By using our proposed scheme, the context-capture in fusion center is shown in Fig. 5. The results show that our scheme can capture efficiently all the local decisions contexts from the sensors.

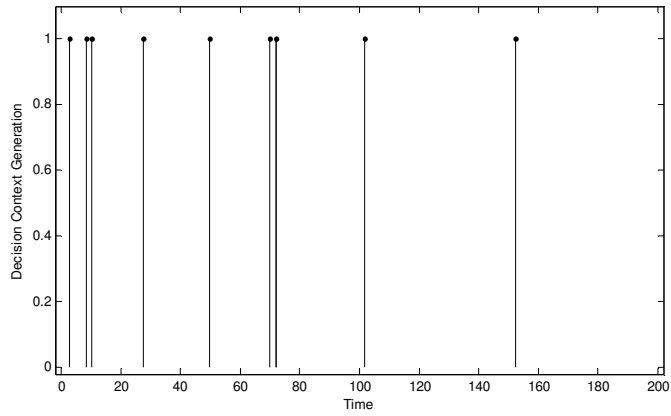

Figure 4.  Local decision context in Sensor-1

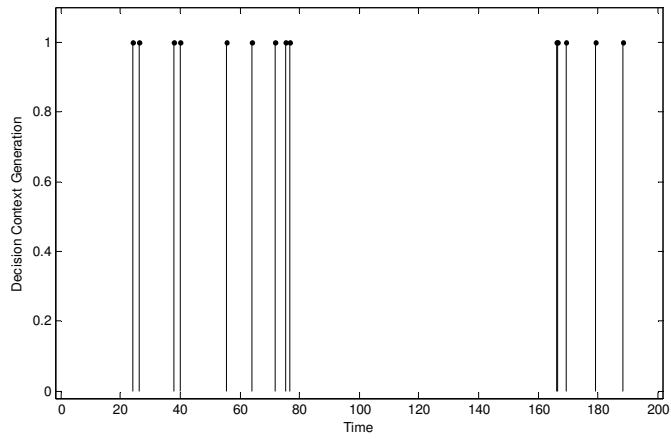

Figure 5.  Local decision context in Sensor-2

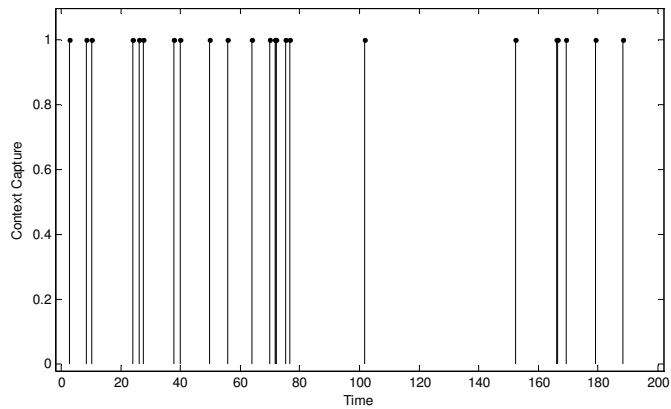

Figure 6.  Context-capture in fusion center

## 7.2. Fault Tolerant

This section we simulate and evaluate the fault tolerant capability. We take ternary logic as the example. In the simulation, there are 5 fault stuck sensor. For evaluating the fault tolerant capability of our scheme, we assume the observation signal-to-noise ratio (OSNR) of each sensor is 2dB. We consider three cases. The first case is normal binary decision fusion without considering fault protection and bursty context-capture. The second case is ternary fault-tolerant decision fusion without considering bursty context-capture. The third case is our multi-valued decision fusion with considering bursity context-capture.

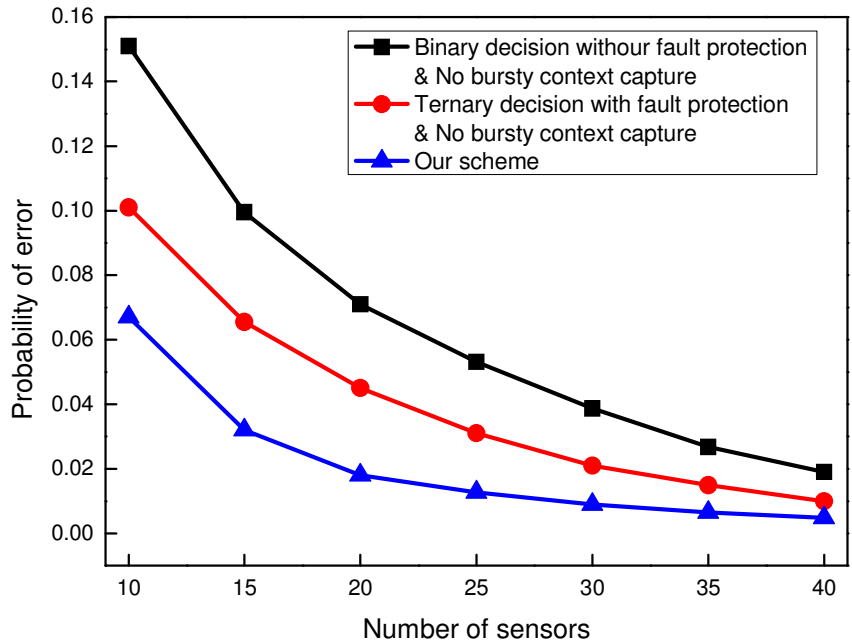

Figure 7. Probability of error

Figure 7 shows the error probability. Note that we use ternary value decision as the example of multi-value decision for evaluation. As shown in Fig. 7, for the same network size, the probability of error of multi-valued decision fusion is lower than that of binary decision fusion. This is because that we introduce the MVL based fault detection into WSNs. Also, the case considering bursty context has lower error probability than the case without considering bursty context. This is because that the scheme with bursty context-capture can enhance the probability of getting more true local decision from sensors. In particular, the fault tolerant capability of our scheme is very obviously when the size of networks is small. In all, our context-capture multi-valued has great advantage on enhance the fault tolerant capability of WSNs.

## 8. CONCLUSIONS

In this paper, we have presented a context-capture multi-valued decision fusion scheme for WSNs. On one hand, MMPP is introduced to deal with bursty characteristic in WSNs. Superposition is used to construct the multi-state of MMPP. On the other hand, MVL is used to perform the multi-valued decision fusion. Fault detection is performed based MVL. Then our scheme removes the fault decision from the decision fusion process, where the fusion rule is based likelihood ratio. Furthermore, through bursty context-capture and multi-valued decision, our scheme can reduce obviously the probability of error during the decision fusion. The simulation results show several advantages of our scheme. Our scheme can capture the bursty context efficiently. Moreover, our scheme has efficient fault tolerant capability. Consequently, the proposed scheme is applicable for decision fusion in WSNs.

## ACKNOWLEDGEMENTS

This work was supported by Japan Society for the Promotion of Science (JSPS) under Grant-in-Aid for Scientific Research(C) (No.20560373) and by the Ph. D. Fellowship program of the China Scholarship Council (No.2008638003).

## REFERENCES

[1]     A. Lei and R. Schober, "Coherent Max–log Decision Fusion in Wireless Sensor Networks," IEEE Transactions on Communications, vol. 58, no. 5, pp. 1327-1332, May 2010.

[2]     C. Suh and Y. B. Ko, "Design and Implementation of Intelligent Home Control Systems based on Active Sensor Networks," IEEE Transactions on Consumer Electronics, vol. 54, no. 3, pp. 1177-1184, Aug. 2008.

[3]     K. Xing, M. Ding, X. Cheng, and S. Rotenstreich, "Safety Warning Based on Highway Sensor Networks," in Proc. of IEEE WCNC 2005, pp. 2355-2361, Mar. 2005.

[4]     S. Schaust, and M. Drozda, "Impact of Packet Injection Models on Misbehaviour Detection Performance in Wireless Sensor Networks," in Proc. of IEEE MASS 2007, Pisa, Italy, Oct. 2007.

[5]     L. Qiao, D. Agrawal, and A. E. Abbadi, "Supporting sliding windows queries for continuous data stream," in Proc. of 15th International Conference on Scientific and Statistical Database Management (SSDBM 2003), MA. USA, Jul. 2003.

[6]     A. Akbari, A. Dana, A. Khademzadeh and N. Beikmahdavi, "Fault detection and recovery in wireless sensor network Using Clustering," International Journal of Wireless & Mobile Networks (IJWMN), vol. 3, no. 1, pp. 130-138, Feb. 2011.

[7]     M. Asim, H. Mokhtar and M. Merabti, "A self-managing fault management mechanism for wireless sensor networks," International Journal of Wireless & Mobile Networks (IJWMN), vol.2, no.4, pp. 184-197, Nov. 2010.

[8]     J. O. Kim, P. Lala, Y. G. Kim, and H. S. Kim, "Fault analysis of the multiple valued logic using spectral method," in Proc. of 30th IEEE International Symposium on Multiple-Valued Logic (ISMVL 2000), May 2000.

[9]     T. Y. Wang, L. Y. Chang, D. R. Duh, and J. Y. Wu, "Fault-tolerant decision fusion via collaborative sensor fault detection in wireless sensor networks," IEEE Transactions on Wireless Communications, vol. 7, no. 2, pp. 756-768, Feb. 2008.

[10]     Q. Cheng, P. K. Varshney, K. G. Mehrotra, and C. K. Mohan, "Bandwidth management in distributed sequential detection," IEEE Transactions on Information Theory, vol. 51, no. 8, pp. 2954–2961.

[11]    S. Marano, V. Matta, P. Willett, and L. Tong, "SPRTs in sensor networks with mobile agents," in Proc. 6th IEEE Workshop on Signal Processing Advances in Wireless Communications, Jun. 2005.

[12]    Q. Zhang and P. K. Varshney, "Decentralized M-ary detection via hierarchical binary decision fusion," Information Fusion, vol. 2, pp. 3-16, Mar. 2001.

[13]    C. Rorres X. Zhu, Y. Yuan, and M.Kam, "M-ary hypothesis testing with binary local decisions," Information Fusion, vol. 5, no. 3, pp. 157-167, Sep. 2004.

[14]    N. Ansari, E. S. H. Hou, B. Zhu, and J. Chen, "Adaptive fusion by reinforcement learning for distributed detection systems," in IEEE Transactions on Aerospace and Electronic Systems, vol. 32, no. 2, pp. 524-531, April 1996.

[15]    X. Zhu, M. Kam, and Q. Zhu, "Adaptive Bayesian decision fusion," in Proc. 1997 IEEE Conference on Decision and Control, San Diego, CA, pp. 5004-5009, December 1997.

[16]    G. Mirjalily, Z. Luo, T. N. Davidson, and É. Bossé, "Blind adaptive decision fusion for distributed detection," IEEE Transactions on Aerospace and Electronic Systems, vol. 39, no. 1, pp. 34-52, Jan. 2003.

[17]    W. Fischer and K. Meier-Hellstern, "The markov-modulated poisson process (MMPP) cookbook," Performance Evaluation, vol. 18, pp. 149-171, 1992.

[18]    Q. Du, "A montonicity result for a single-server queue subject to a Markov-modulated Poisson process," Journal of Applied Probability, vol. 32, pp. 1103-1111, 1995.

[19]    M. Yu and D. G. Daut, "A new traffic aggregation technique based on Markov modulated Poisson process," in Proc. of IEEE GLOBECOM 2005, Nov./Dec. 2005.

[20]    P. Salvador and R. Valadas, "A framework based on Markov modulated Poisson Process for modeling traffic with long-range dependence," In Proc. of SPIE Conference on Internet Performance and Control of Network Systems II, Denver, Colorado, USA, Aug. 2001.

[21]    G. Epstein, G. Frieder, and D.C. Rine, "The Development of Multiple-Valued Logic as Related to Computer Science," Computer, vol. 7, pp. 20-32, 1974.

[22]    S.L. Hurst, "Multiple-Valued Logic—Its Status and Its Future,"IEEE Transactions on Computers, vol. 33, no. 12, pp. 1,160-1,179, Dec. 1984.

[23]    K.C. Smith, "A Multiple-Valued Logic: A Tutorial and Appreciation," Computer, vol. 21, no. 4, pp. 17-27, Apr. 1988.

[24]    B. Devereux and M. Chechik, "Edge-Shifted Decision Diagrams for Multiple-Valued Logic." In JMVLSC, Old City Publishing, 2003.

[25]    S. Scott, "Bayesian methods and extensions for the two state Markov modulated Poisson process," Ph. D. thesis, Harvard University, Department of Statistics, 1998.

[26]    M. F. Neuts, "Structured stochastic matrices of M/G/1 type and their applications," Probability: Pure and Applied, vol. 5, New York: Marcel Dekker, 1989.

[27]    M. G. Karpovsky, Spectral Techniques and Fault Detection, New York, Academic Press, 1985.

## Authors

**Jun Wu** was born in Hunan, China. He is currently pursuing the Ph.D. degree at Waseda University, Japan, funded by China Scholarship Council. He received the B.S. degree and M.S. degree from the Information Engineering College, Xiangtan University, Hunan, China, in 2002 and 2005, respectively. From July 2005 to September 2008, he was a lecturer in the Xiangtan University. His current research interests include wireless sensor networks and their information security. Mr. Wu is a student member of the IEEE.

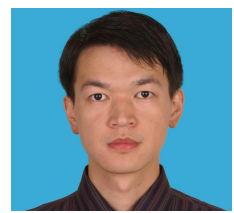

**Shigeru Shimamoto** was born was born in Mie, Japan, in 1963. He received the B.E and M.E. degrees in communications engineering from the University of Electro Communication, Tokyo, Japan, in 1985 and 1987, respectively. He received the Ph. D. degree from Tohoku University, Japan in 1992. From April 1987 to March 1991, he joined NEC Corporation. From April 1991 to September 1992, he was an Assistant Professor in the University of Electro Communications, Tokyo, Japan. He has been an Assistant Professor in the Gunma University from October 1992 to December 1993. Since January 1994 to March 2000, he has been an Associate Professor in Department of Computer Science, Faculty of Engineering, Gunma University, Gunma, Japan. Since April 2002, he has been a Professor at GITS, Waseda University. In 2008, he also served as a visiting professor at Stanford University, USA. His main fields of research interest include sensor networks, satellite and mobile communications, optical wireless, Ad-hoc networks and body area network. Dr. Shimamoto is a member of the IEEE.

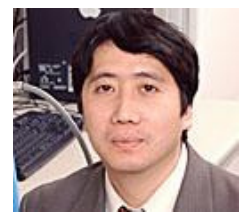